\documentclass[aps,prc,nofootinbib,tightenlines,notitlepage, twocolumn,floatfix,superscriptaddress,showkeys]{revtex4-1}
    
\usepackage[utf8]{inputenc}          
\usepackage{graphicx}         
\usepackage[usenames,dvipsnames]{xcolor}
\usepackage{array,dcolumn,longtable} 
\usepackage{gensymb}

\usepackage{enumerate}

\usepackage{amsmath,amssymb,amsfonts,slashed} 

\usepackage{hyperref}
\hypersetup{linktocpage,breaklinks}

\usepackage{txfonts}
\usepackage{bm}
\usepackage{stmaryrd}
\usepackage[utf8]{inputenc}

\usepackage{epsfig}
\usepackage{epstopdf}

\usepackage{natbib}
\usepackage{cleveref}

\definecolor{mred}{RGB}{127,0,25}
\definecolor{mdgr}{RGB}{51,51,51}
\definecolor{mag}{RGB}{211, 54, 130}
\definecolor{verm}{RGB}{164, 25, 0}

\hypersetup{colorlinks=true,
            citecolor=NavyBlue,
            linkcolor=NavyBlue,
            urlcolor=NavyBlue}

\usepackage{xcolor}

\newcommand{\beq}{\begin{equation}}
\newcommand{\eeq}{\end{equation}}
\newcommand{\bea}{\begin{eqnarray}}
\newcommand{\eea}{\end{eqnarray}}

\newcommand{\calO}{{\cal O}}

\begin{document}

\preprint{APS/123-QED}

\title{Mapping out the thermodynamic stability of a QCD equation of state \\ with a critical point using active learning}

\author{D. Mroczek}
\affiliation{Illinois Center for Advanced Studies of the Universe, Department of Physics, University of Illinois at Urbana-Champaign, Urbana, IL 61801, USA}
\author{M. Hjorth-Jensen}
\affiliation{Department of Physics and Astronomy and Facility for Rare Ion Beams, Michigan State University, MI 48824, USA, and Department of Physics and Center for Computing in Science Education, University of Oslo, N-0316 Oslo, Norway}

\author{J. Noronha-Hostler}%
\affiliation{Illinois Center for Advanced Studies of the Universe, Department of Physics, University of Illinois at Urbana-Champaign, Urbana, IL 61801, USA}%


\author{P. Parotto}
\affiliation{
Pennsylvania State University, Department of Physics, University Park, PA 16802, USA
}%
\author{C. Ratti}
\affiliation{Department of Physics, University of Houston, Houston, TX 77204, USA}%

\author{R. Vilalta}
\affiliation{%
 Department of Computer Science, University of Houston, Houston, TX 77204, USA}%

\date{\today}

\begin{abstract}The Beam Energy Scan Theory (BEST) collaboration’s equation of state (EoS) incorporates a 3D Ising model critical point into the Quantum Chromodynamics (QCD) equation of state from lattice simulations. However, it contains 4 free parameters related to the size and location of the critical region in the QCD phase diagram. Certain combinations of the free parameters lead to acausal or unstable realizations of the EoS that should not be considered. In this work, we use an active learning framework to rule out pathological EoS efficiently. We find that checking stability and causality for a small portion of the parameters' range is sufficient to construct algorithms that perform with $>$96\% accuracy across the entire parameter space. Though in this work we focus on a specific case, our approach can be generalized to any EoS containing a parameter space--class correspondence.

\end{abstract}

\maketitle


\section{\label{sec:intro}Introduction}

    

The phase diagram of nuclear matter, while widely studied, remains mostly unknown. It has been established from lattice simulations of Quantum Chromodynamics (QCD) that the transition from hadronic to quark degrees of freedom is a crossover at vanishing baryon chemical potential~\cite{Aoki:2006we}. Effective models predict this transition will become first-order at finite densities (for a review see Refs. \cite{Stephanov:2004wx,Stephanov:1998dy}). Since large-scale, first-principle lattice QCD calculations cannot yet be performed directly at finite baryon density, experimental searches for the critical point (CP) and a first-order phase transition are vital in determining the phase structure of QCD at different densities. Preliminary results from the first phase of the Beam Energy Scan (BES-I) program at the Relativistic Heavy-Ion Collider (RHIC) showed promising trends in the data \cite{STAR:2020tga,STAR:2021fge,STAR:2020dav}. These will be confirmed or disproved  during the second phase of the program, BES-II, which ran through 2021 with improvements to detectors and statistics. The determination of the phase structure of QCD, along with the existence and location of its critical point, remains among the most important goals of high-energy nuclear physics in view of results from BES-II \cite{Aprahamian:2015qub,Bzdak:2019pkr,Dexheimer:2020zzs,Monnai:2021kgu,An:2021wof}. At even lower beam energies, the HADES (High Acceptance Di-Electron Spectrometer) experiment is searching for a first-order phase transition to support the presence of a critical point \cite{HADES:2020wpc}. 

 Previously, a key factor limiting research of critical signatures on the theoretical side was the lack of an equation of state (EoS) including a critical point in the correct universality class and matching what is already known from lattice simulations. Such an EoS is now available and ready to be implemented in hydrodynamic simulations at BES-II energies \cite{Parotto:2018pwx,Karthein:2021nxe}. Results from such simulations are essential for the analysis of BES-II measurements, because they can provide precise calculations of higher order net-proton cumulants as functions of the collision energy $\sqrt{s_{NN}}$ -- promising experimental signatures for criticality \cite{Pradeep:2021opj,An:2021wof,An:2020vri,Nahrgang:2018afz}. Moreover, they could help quantify the likelihood that such signatures survive final hadronic scatterings. A first effort in that direction was presented in Ref. \cite{Mroczek:2020rpm}, where the effects of a critical point on the fourth order baryon number susceptibility $\chi^B_4$, accessible experimentally via net-proton kurtosis measurements, were studied in the context of the parameterized EoS introduced in Ref. \cite{Parotto:2018pwx}. Much work needs to be done before direct theory-to-experiment comparisons can be made, including adjustments in hydrodynamic calculations near the critical point \cite{Stephanov:2009ra, Nahrgang:2011mg, Stephanov:2017ghc, Nahrgang:2018afz,Bluhm:2020mpc}. Once these modifications are quantified, the EoS in Ref. \cite{Parotto:2018pwx} would allow for a precise survey of collisions at BES-II energies. 

The procedure described in Ref. \cite{Parotto:2018pwx} is based on combining a critical point in the 3D Ising model universality class and lattice QCD results in the form of a Taylor expansion. This requires the Ising variables  $(r,h)$ -- reduced temperature $r$ and magnetic field $h$, respectively -- to be mapped to QCD variables $(T,\mu_B)$ -- temperature and baryon chemical potential, although the nature of this mapping is not fixed from first-principles. One is free to choose a map, which might lead to a particular parameterization of the EoS that is not thermodynamically stable and causal by construction. Therefore, a filtering of viable equations of state must take place post EoS computation. In order to do this, a number of thermodynamic quantities must be calculated across the phase diagram, and then thermodynamics inequalities must be verified for all $(T,\mu_B)$. A machine-learning (ML) assisted classification would clearly provide a computational advantage if the process of computing and checking multiple quantities over a grid could be eliminated. The choice to eliminate these steps is motivated by the fact that all thermodynamic quantities relevant for stability analyses are directly related to derivatives of the pressure. Therefore, all the information needed is encoded in the pressure itself. Moreover, for the EoS presented in Ref.~\cite{Parotto:2018pwx}, once the lattice input is chosen, the properties of the EoS are dictated solely by the input parameters, which can in turn be mapped to stable and causal, acausal, and unstable realizations of the EoS. 
This second option, using input parameters instead of the pressure to determine stability and causality, allows one to bypass the computation of the EoS entirely once a ML model learns the acceptable parameter space regions. 

In addition to using traditional supervised learning to tackle the EoS stability and causality problem, we incorporate active learning \cite{settles2009} into the training pipeline, and compare the performance of models trained in different frameworks. 
In active learning, ML models place queries and request labels for points that are considered most informative; this leads to a significant reduction in the class labeling cost (normally as a logarithmic factor), and a speed up in training, as queries are likely to be placed over points located close to the decision boundary separating examples of different classes. For a review on active learning and query strategies see Refs.~\cite{settles2009active,Hino2020,Pengzhen21}. In this work, we implement active learning to expedite learning the map between the parameter space and output class. The problem of determining the acceptable parameter space range for a high-dimensional model using active learning has been shown to work effectively in Ref. \cite{Caron:2019xkx}. 

With the goal of developing a tool that can quickly rule out pathological EoS, we train a set of classifiers to identify thermodynamically stable and causal realizations of the EoS. We start by defining two viable options for the training data -- one consisting of the set of input parameters, and the other being the pressure as a function of temperature and baryon chemical potential. We then select a set of competitive learning algorithms -- random forests (RF), $K$-nearest-neighbor (KNN), and support vector machines (SVM) -- and train them on the EoS input parameters using both active learning and random sampling \cite{Hastie09,Murphy2012,Bishop2006}. We find that active learning \cite{settles2009active} outperforms random sampling in every case, but the random forests model is the only one that converges to high accuracy within the number of training samples generated. We then train a different random forests classifier using a dimension-reduced version of the pressure -- rather than the EoS input parameters -- as training data, using both active learning and random sampling. The random forests classifier trained on the pressure data using active learning converges even faster and to higher accuracy than the previous random forests model. This systematic exploration of learning and sampling models yields a particular combination that is optimal for the task of thermodynamic stability classification of the EoS model. Lastly, we demonstrate how one of the top performing classifiers can be used to map the stable and causal regions of this particular EoS formulation, which can in turn inform theoretical and experimental studies of the QCD critical point. 

We note that various machine learning algorithms have been previously used in the context of heavy-ion collisions \cite{Pang:2016vdc,Haake:2017dpr,Bielcikova:2020mxw,Steinheimer:2019iso,Du:2020pmp,Boehnlein:2021eym,Mallick:2021wop,Lai:2021ckt}, but we are not aware of other works that have been used to constrain the parameter space of possible critical points through thermodynamic stability. Previous well-known examples are an attempt to identify new signatures of the QCD phase transition \cite{Pang:2016vdc,Steinheimer:2019iso} or classify  jets origination from quarks vs. gluons \cite{Chien:2018dfn,Lai:2021ckt}. We are also not aware of previous investigations in the context of heavy-ion collisions that have used active learning (though it has been employed in other contexts in nuclear theory \cite{Sarkar:2021fpz} and high-energy physics \cite{Buhmann:2021caf,Rocamonde:2022gyw,Caron:2019xkx}). For a recent review of artificial intelligence and machine learning applications in nuclear physics, see Ref.~\cite{Boehnlein:2021eym}.

The paper is structured as follows. Section II summarizes the methodology introduced in~\cite{Parotto:2018pwx} for generating a realization of the EoS. In Section III, we discuss how thermodynamic stability and causality issues can arise in the EoS formulation, the possible formats of the training data, as well as the preprocessing framework. In Section IV, we outline the basic ideas behind active learning and our query strategy. Section V deals with the implementation of our training and sampling methods in the development of the classifiers. Results and conclusions follow.

\section{parameterized EoS with a critical point}\label{EoS}

Due to the fermion sign problem, direct lattice simulations at finite chemical potentials are not possible at the moment. The most straightforward way to work around this problem is to define a Taylor expansion around $\mu_B=0$. For the equation of state, this commonly consists of an expansion of the pressure as:
\begin{equation}
\frac{P}{T^4} (T,\mu_B) = \sum_{n} c_n (T) \left( \frac{\mu_B}{T} \right)^n
\end{equation}
where the coefficients are related to the derivatives of the pressure with respect to the chemical potential:
\begin{equation}
c_n (T) = \frac{1}{n!} \chi^B_n (T) = \frac{1}{n!} \frac{\partial^n (P/T^4)}{\partial (\mu_B/T)^n} \, \, .
\end{equation}

The BEST Collaboration's family of EoS of Ref.~\cite{Parotto:2018pwx} was constructed by incorporating a critical point from the 3D Ising model universality class, and imposing exact matching with lattice QCD results at $\mu_B=0$ (up to order $\calO (\mu_B^4)$). 

We summarize here the steps followed in Ref.~\cite{Parotto:2018pwx} for generating each EoS:
\begin{enumerate}[i)]
\item Define a parameterization of the 3D Ising model EoS in the vicinity of the critical point. This parameterization
imposes the correct critical behavior by expressing the magnetization $M$, the magnetic field $h$ and the reduced temperature $r = (T - T_c) / T_c$, where $T_c$ is the critical temperature,  in terms of new parameters 
$(R,\theta)$ with \cite{Nonaka:2004pg,Guida:1996ep,Schofield:1969zz,Bluhm:2006av}: 
\begin{align} \label{eq:param_Ising} \nonumber
M &= M_0 R^\beta \theta \, \, , \\
h &= h_0 R^{\beta \delta} \tilde{h}(\theta) \, \, , \\ \nonumber 
r &= R (1 - \theta^2) \, \, ,
\end{align}
where $M_0 \simeq 0.605$ and $h_0 \simeq 0.364$ are normalization constants, 
$\tilde{h} (\theta) = \theta (1 + a \theta^2 + b \theta^4)$, with $a= - 0.76201$ and 
$b=0.00804$, and $\beta \simeq 0.326$, $\delta \simeq 4.80$ are 3D Ising model critical 
exponents \cite{Guida:1996ep}. The parameters satisfy $R \geq 0$ and 
$\left| \theta \right| \leq \theta_0$, with $\theta_0 \simeq 1.154$.

\item Map the phase diagram of the 3D Ising model onto that of QCD, in a way that allows one to choose the location of the critical point. This mapping can be done using a simple linear map, which requires six parameters \cite{Rehr:1973zz}:
\begin{align}
\frac{T - T_C}{T_C} &=  w \left( r \rho \,  \sin \alpha_1  + h \, \sin \alpha_2 \right) \, \, , \label{eq:IsQCDmap1} \\ 
\frac{\mu_B - \mu_{BC}}{T_C} &=  w \left( - r \rho \, \cos \alpha_1 - h \, \cos \alpha_2 \right) \, \, , \label{eq:IsQCDmap2}
\end{align} 
where $(T_C,\mu_{BC})$ indicate the location of the critical point, while $(\alpha_1, \alpha_2)$ are the angles between the horizontal $(T=\textit{const})$ lines on the QCD phase diagram and the $h=0$ and $t=0$ Ising model axes, respectively. The size of the critical region is roughly determined by the scaling parameters $w, \rho$ in the Ising-to-QCD map~\cite{Pradeep:2019ccv,Mroczek:2020rpm}.

The number of free parameters is reduced from six to four by imposing that the critical point is located on the chiral transition line predicted by lattice QCD:
\begin{equation}\label{eq:trline}
T = T_0 + \kappa_2 \, T_0 \left( \frac{\mu_B}{T_0} \right)^2 + {\cal O} (\mu_B^4),
\end{equation}
which fixes the value of $T_C$ and $\alpha_1$, given a choice of $\mu_{BC}$.

The lattice QCD input for the pressure and its derivatives at $\mu_B = 0$ is from the Wuppertal-Budapest Collaboration~\cite{Borsanyi:2013bia, Bellwied:2015lba}, and the QCD transition line is assumed to be a parabola with curvature $\kappa_2 = -0.0149$, as estimated in Ref.~\cite{Bellwied:2015rza}. This is a valid assumption in the range of chemical potentials covered by the BEST EoS and the BES-II program. Although more recent results have since become available, determining the ``hyper-curvature" $\kappa_4$~\cite{HotQCD:2018pds,Borsanyi:2020fev} of the transition line, it is found to be consistent with zero within errors.

\item Impose that the EoS exactly matches lattice QCD at $\mu_B=0$ by requiring that the expansion coefficients determined from the lattice are a sum of a contribution from the critical point, and a ``regular" one
\begin{equation} \label{eq:coeffs}
T^4 c_n^{\text{LAT}} (T) = T^4 c_n^{\text{Non-Ising}} (T) + T_C^4 c_n^{\text{Ising}} (T) \, \, ,
\end{equation}
where $c_n^{\text{LAT}}$ are the coefficients calculated from the lattice, and 
$c_n^{\text{Ising}}$ determine the contribution from the critical point. The coefficients $c_n^{\text{Non-Ising}}$ contain the contribution to the thermodynamics at 
$\mu_B=0$ not due to the critical point. The procedure is carried out up to order $\calO (\mu_B^4)$.

\item Reconstruct the full QCD pressure as the sum of the ``Ising" and ``Non-Ising" contributions
\begin{equation} \label{eq:Pfull}
P (T, \mu_B) = T^4 \sum_n c_n^{\text{Non-Ising}} (T) \left( \frac{\mu_B}{T} \right)^n + P^{\text{QCD}}_{\text{crit}}(T, \mu_B) \, \, ,
\end{equation}
where $P^{\text{QCD}}_{\text{crit}}(T, \mu_B)$ is the critical pressure mapped onto QCD 
from the 3D Ising model. For additional details we refer to Ref.~\cite{Parotto:2018pwx}.
\end{enumerate}

With the construction just summarized, the pressure in Eq.~\eqref{eq:Pfull} only depends on the non-universal mapping between the 3D Ising model and QCD, which is ultimately fixed by the parameters $\mu_{BC}$, $\alpha_{\rm diff} = \alpha_2 - \alpha_1$, w, and $\rho$. The complete thermodynamic description is in turn obtained by computing the baryon number density 
\begin{equation}
      \dfrac{n_B(T,\mu_B)}{T^3} = \dfrac{1}{T^3}\left(\dfrac{\partial P}{\partial \mu_B}\right)_T \, \, ,
\end{equation} 
entropy density 
\begin{equation}
    \dfrac{s (T,\mu_B)}{T^3} = \dfrac{1}{T^3}\left(\dfrac{\partial P}{\partial T}\right)_{\mu_B} \, \, ,
\end{equation}
energy density 
\begin{equation}
 \dfrac{\varepsilon(T,\mu_B)}{T^4} = \dfrac{s}{T^3} - \dfrac{P}{T^4} + \dfrac{\mu_B}{T}\dfrac{n_B}{T^3}    \, \, ,
\end{equation}
and speed of sound 
\begin{equation}
c^2_s (T,\mu_B) = \left(\dfrac{\partial P}{\partial \varepsilon}\right)_{s/n_B} \, \, ,
\end{equation}
all of which are normalized by the correct power of the temperature.

More recently, this formulation has been updated to account for strangeness neutrality, which is relevant in heavy-ion collisions~\cite{Karthein:2021nxe}. In this work, we implement the original description of the EoS, assuming vanishing net strangeness and electric charge chemical potentials, $\mu_S = \mu_Q = 0$.

\section{Training}\label{sec:training}

Developing a successful classifier requires a thoughtful selection of training data. We can think of the EoS framework as a set of two maps:
\begin{equation}\label{eq:maps}
    (\mu_{BC}, \alpha_\textrm{diff}, w, \rho) \mapsto P(T,\mu_B) \mapsto \{\textrm{acceptable, unstable, acausal}\}.
\end{equation}

\begin{figure*}
   \centering
\begin{tabular}{c}
\includegraphics[width=.9\linewidth]{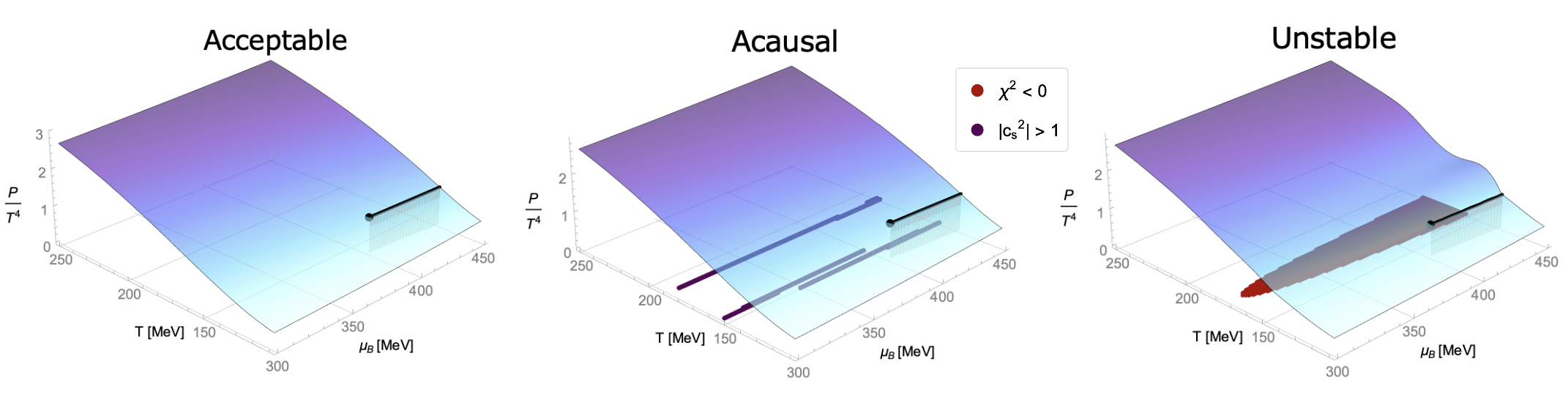}
\end{tabular}

    \caption{Different realizations of the EoS pertaining to the three thermodynamic stability classes. The black line illustrates the first-order transition line, which ends at the critical point. The projected $T-\mu_B$ plane is also shown, on which the points where some pathological behavior appears are highlighted (purple corresponding to causality violations and red to negativity of the second-order baryon susceptibility).}
    \label{fig:EoS}
\end{figure*}

The first map yields the pressure as a function of temperature and chemical potential. The second map determines if the resulting EoS is acceptable or not. Fig.~\ref{fig:EoS} illustrates how different realizations of the EoS can present pathological behaviors.

We use both the input parameters and a dimension-reduced version of the pressure for training. It is important to investigate how  the choice of training -- on either the input parameters or pressure -- map to stable EoS, because the two spaces relate to thermodynamic stability in fundamentally different ways.
The key difference is that models trained on input parameters are constrained to this particular formulation of the EoS, while training on the pressure can yield ML models that generalize to any QCD EoS. However, we will not test how training on the pressure can be applied to alternative EoS since that is beyond the scope of this paper. In this work, we focus on establishing that it is possible to train ML classifiers that identify viable EoS quickly and with high accuracy. Below we discuss how the labels for the training set were created and how the data was processed prior to training.

\subsection{Thermodynamic Stability}

There are four free parameters that emerge from our construction of the EoS that can lead to pathological behavior. Hence, thermodynamic stability needs to be verified at the end of the procedure, once all thermodynamic quantities have been calculated. In general, we require the positivity of the pressure, entropy and baryon density, the second order baryon susceptibility ($\chi_2^B$), and the heat capacity $(\partial S / \partial T)_{n_B}$  \begin{align}\label{eqn:stabcon}
    & P, s, \varepsilon, n_B, \chi_2^B, \left(\frac{\partial S}{\partial T}\right)_{n_B} > 0 \, \, ,
\end{align}
which follows from the requirement that entropy should be maximized in equilibrium.
We also require that the speed of sound squared must be both positive and bounded by causality
\begin{align}\label{eqn:cs2con}
    & 0 \leq c^2_s \leq 1.
\end{align}

All of these must be satisfied at every point in the $(T-\mu_B)$ plane. As mentioned in Section~\ref{sec:intro}, there are two options for training data -- using the pressure, which encodes all the information regarding stability and causality, or the input parameters which map to a particular EoS. In both cases, some number of training samples need to be generated. These must go through the numerical differentiation and grid checking process to be labeled. Generating a single EoS with a complete thermodynamic description and label can take between 1.2-4 times as long as generating the pressure only. While this approach reduces computation time and memory requirements,
the difference is even more dramatic if we use the input parameters as training data. No calculation is required for generating an unlabeled sample of the parameter space -- we simply select points from the parameter space grid. The runtime is negligible compared to how long it takes to generate a realization of the EoS.

\subsection{Preprocessing}\label{preprocessing}

Let us first discuss the dimensionality of our problem at each stage of the mapping described in Eq.\ (\ref{eq:maps}). 
We define a single set of input parameters as a training vector $\vec{\Omega}_{(i)} = (\mu_{B (i)}, \alpha_\textrm{diff(i)}, w_{(i)}, \rho_{(i)})$. Each $\vec{\Omega}_{(i)}$ is four-dimensional, so we do not need to apply dimensionality reduction techniques. The only pre-processing required is a standard scaling of the distribution of $\vec{\Omega}_{(i)}$ in the training set. This is necessary to ensure the different features are compared along the same scale, to avoid artificially introducing differences in the data, and to ensure the pool and test set are analyzed with respect to the training distribution. We discuss the role of the pool, training, and test sets in the next section.

In the next step of our mapping, $P(T,\mu_B)$ has dimensions corresponding to the grid size of the EoS. In our case, the limits are
\begin{eqnarray}
30 \leq \, &T\; [\rm MeV]& \, \leq 800\\
0 \leq \, &{\mu_B}\; [\rm MeV]& \, \leq 450
\end{eqnarray}
with a step size of $1$ MeV in both directions. 
Thus, $P(T,\mu_B)$ is a table of dimension $451\times771$. A grid of this magnitude is not optimal for machine learning. We use the standard technique of Principal Component Analysis (PCA) \cite{Jolliffe2002,Hastie09,Murphy2012} to create a dimension-reduced projection of the original matrix defined by $P(T,\mu_B)$. We check how many components are needed to account for most of the variance present in the pressure; our results show that the two-component projected matrix accounts for over 99\% of the variance in nearly all cases. Based on these findings, we define a new variable, $P^*$, with about 1500 features corresponding to the two columns of the two-dimension projection matrix from the PCA. This is by no means a low-dimension feature space, but it is easily handled by most machine learning algorithms.  

The final stage of the learning pipeline classifies the input EoS as either acceptable, unstable and acausal, or acausal, with no signs of instability (3-dimensional output space). 

\section{Sampling}\label{sec:sampling}

Since the parameter space for our model is continuous, we first discretize it by defining a grid in each parameter within a range of interest. The bounds and step-sizes for each parameter are summarized in Table ~\ref{tab:params}.
The bounds in $\mu_{BC}$ are motivated by lattice QCD constraints. The upper bound is informed by the fourth order Taylor expansion of lattice data used in Ref.~\cite{Parotto:2018pwx}, which breaks down at $\mu_{B} \gtrapprox 450 $ MeV. While there is no limit to how close the critical point can be placed to vanishing chemical potentials by construction, lattice results indicate that the region $\mu_B \lessapprox 2T$ is not likely to contain a critical point \cite{Bazavov:2017dus}. The lower bound in $\mu_{BC}$ is loosely determined by these results and based on lattice calculations for the crossover temperature $T_0 \simeq 155$ MeV at $\mu_B = 0$ \cite{Aoki:2009sc,Borsanyi:2010bp,Bhattacharya:2014ara,Bazavov:2011nk}. Since the curvature of the deconfinement transition line appears to be negative, we can safely expect that the critical temperature $T_C \lessapprox 155$ MeV. Lattice results then roughly rule out $\mu_{BC} \lessapprox 300$ MeV, but we extend this lower bound down to $220$ MeV to accommodate for possible uncertainties in these values. The other three parameters are specific to the linear map assumed in the construction of the EoS and no arguments from first-principles constrain their values, so the corresponding bounds are designed to span all possible behavior. Broadly speaking, it was observed already in Ref.~\citep{Parotto:2018pwx} that, with all other parameters fixed, when a certain choice of $w$ was found to be pathological, then the same occurred for all $w'<w$. The opposite behavior was observed for a pathological EoS with a certain $\rho$ --  with all other parameters fixed, all $\rho'>\rho$ were pathological.

Every time a model is initialized, an initial training and test sets are generated from the parameter grids, containing 350 and $\sim20,000$ labeled realizations of the EoS, respectively. The initial training set is chosen randomly at the beginning of each training cycle from the remaining points (which constitute the pool set), so that there is no overlap between test and training sets. The initial training set contains the first labeled realizations from which the model will learn. More instances are added to the training set with each iteration. The test set remains the same throughout all training iterations and it is used to check the accuracy of the model at each stage. 

Once the initial training set $\mathcal{L}_0$ has been determined, we take the following steps:
\begin{enumerate}
     
    \item A machine learning model is trained on $\mathcal{L}_0$.
    \item The model makes a prediction on the test set and its performance is recorded (for reporting purposes only).
    \item The model is then evaluated on $\mathcal{U}_0$, the pool set, which contains all points in neither $\mathcal{L}_0$ nor the test set. These points are unlabeled.
    \item Using some selection criterion, a query is generated. This means a size-$k$ pool of points is selected from $\mathcal{U}_0$ according to some distribution, and a label is provided for these points. 
    \item The training and pool sets are updated. The new training set $\mathcal{L}_1$ contains $\mathcal{L}_0$ and the queried points, which are now missing from the updated pool set $\mathcal{U}_1$.
    \item The model is trained on $\mathcal{L}_1$ and steps 2-5 are repeated until a stopping criterion is met.
\end{enumerate}

\begin{table}[htb]
  \begin{center}
    \caption{Ranges and step-sizes used to generate EoS for training and testing.}
    \label{tab:params}
\begin{tabular}{|c|ccc|}
\hline
 & Min. & Max. & Step size \\
\hline
 $\mu_{BC}$ & 220 MeV & 420 MeV & 20 MeV \\
 $w$ & 0.1 & 10.0 & 0.5 \\
 $\rho$ & 0.1 & 10.0 & 0.5 \\
 $\alpha_{\textrm{diff}}$ & $-180\degree$ & $180\degree$ & $5\degree$ \\
\hline
    \end{tabular}
  \end{center}
\end{table}

We expect to see the average recorded accuracy increase as sample size increases until the model converges at some maximum accuracy value, or until resources for generating labeled instances have been exhausted. This should happen independently of the selection criterion. However, if labeled samples are difficult (e.g. computationally costly) to generate, an improved sampling method can provide an advantage in terms of how many samples are needed to achieve a target performance. Active learning methods seek to increase the performance of learning algorithms with fewer samples by allowing models to choose which data to learn from. 

We test the performance of our models using both random sampling and active learning. We draw our samples in a pool-based fashion, meaning queries consist of size-$k$ samples drawn from $\mathcal{U}$. 
In the random case, the samples are randomly pulled from $\mathcal{U}$ assuming a uniform distribution. Margin-based queries select the k points currently in the pool set $\mathcal{U}$ with the smallest margin values, where the margin $M$ is defined as in Ref.~\cite{scheffer2001active},
\begin{equation}
        M = P(\hat{y}_1) - P(\hat{y}_2)
\end{equation}
 and $\hat{y}_1$ and $\hat{y}_2$ are the first and second most probable class labels under the current model, with corresponding probabilities $P(\hat{y}_i)$.  Therefore, this sampling method favors points with a small margin, meaning the classification is ambiguous, whereas points where one class is clearly preferred do not get labeled. Using this query strategy avoids wasting resources on instances the model already understands how to classify in favor of those that are still ambiguous.  It is important to note that, although we make the choice to fix each query at $k = 200$ samples  (i.e. each iteration in training represents the same increase in training set size) that choice is in principle arbitrary. 

\section{Model training and selection}

The main goal in the model training and selection stage is to gauge what is necessary to create a strong EoS classifier -- how much data is needed, how to sample from the available data, and how to make a choice for the classification algorithm. 
Generally, the amount of data needed is measured according to the accuracy of the classifier on the test set, but it could also be limited by computational resources. The preference for a sampling method is determined based on whether random sampling or active learning reached higher accuracy at a lower number of samples (e.g. 95\% test set accuracy rate at 5,000 samples is better than a 95\% rate at 7,000 samples). The best model is then the combination of algorithm plus sampling method that reaches the highest accuracy rate with the fewest possible samples. 



\begin{figure*}[htb] 
\includegraphics[width=0.8\textwidth]{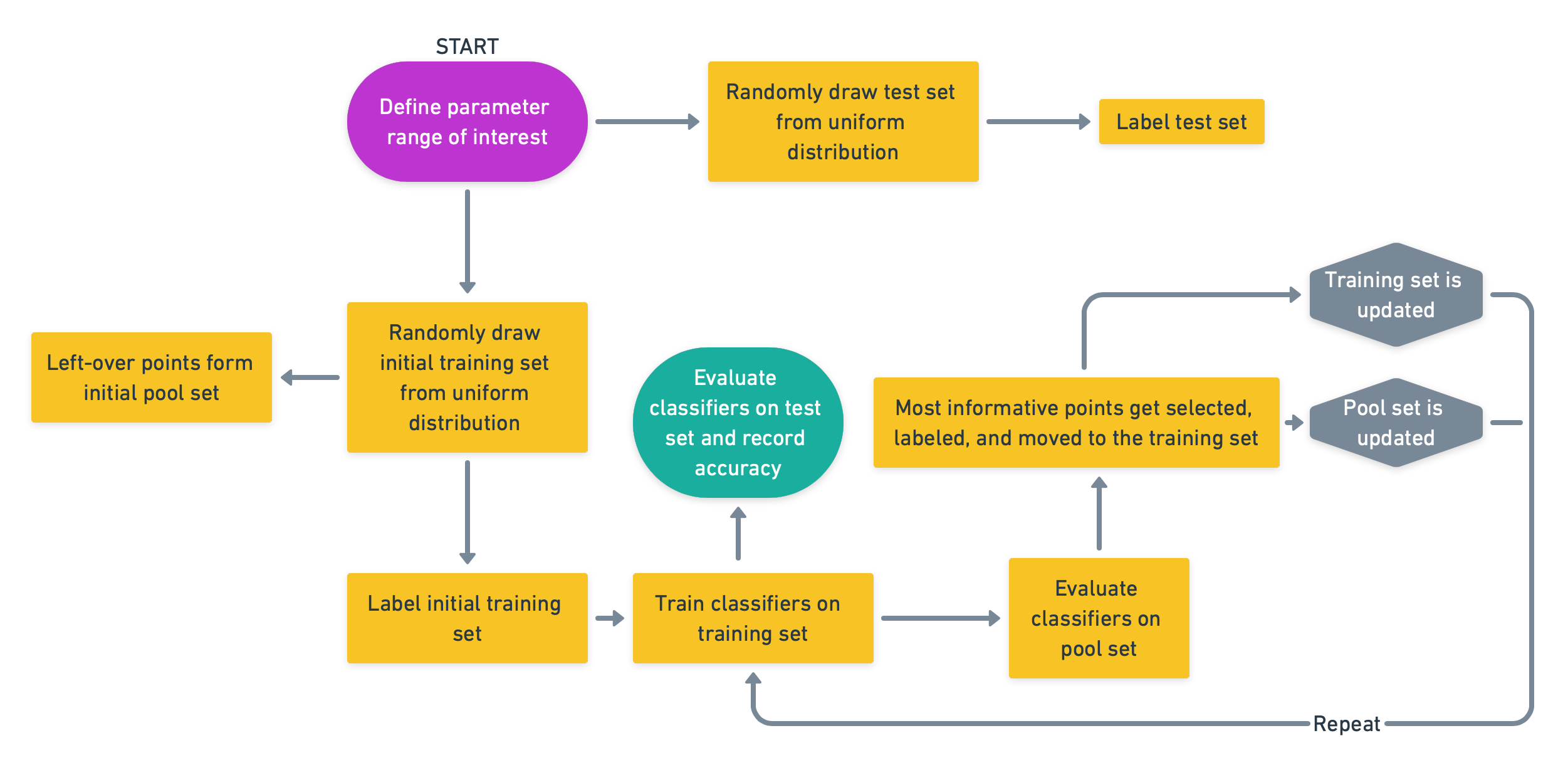}
\caption{Flowchart of model development and testing. From an initial grid of interest, three sets are created -- an initial training set and a test set (each containing labeled realizations of the EoS), and a pool set, which contains all the points outside the test or initial training sets. With each iteration the performance of the classifier is recorded, and a new set of points from the pool set is labeled and moved to the training set.} \label{fig:flowchart}
\end{figure*}

We select three classification algorithms as mentioned in Sec.~\ref{sec:intro} -- SVM, RF, and KNN -- and train them using the sampling framework described in Sec.~\ref{sec:sampling}. The sampling and training procedures are summarized in Fig.~\ref{fig:flowchart}. We used the open source library scikit-learn \cite{scikit-learn} and the publicly available implementation in Ref. \cite{cohen2018activelearningtutorial} to develop the code used in this work. Sampling is performed as outlined in Section~\ref{sec:sampling}. Additionally, at each training step, the model's hyperparameters are optimized over a random grid search using 5-fold cross-validation on the training set. This step is crucial since the training set changes with each iteration and the hyperparameters need to be readjusted. Details about the hyperparameters for each model can be found in the documentation for Ref.~\cite{scikit-learn}, and the specific grid search used in this work is available in the source code \cite{mroczekgithub2022}.

For RF and KNN methods, training is set to stop at 10,000 samples, regardless of accuracy levels, in order to constrain computational expenses. For SVM models, the run-time scales with the cubic power of the number of training samples, and training is set to stop at 2,500 samples instead. To deal with the cold start problem we randomly select a new initial training set with each new run. The cold start problem refers to the expected model instability when faced with data scarcity, which is common when using active learning on a small sample \cite{Yuan20,Grimova18}. We also do not throw away any labels during training. Once a point is labeled, the label is kept and recycled if the same point is called again by the sampling algorithm in a different run. 

We perform a total of 25 experiments -- 5 repetitions for each of the learning algorithms (RF, SVM, KNN), using either random sampling or active learning on the input parameter vector ${\vec{\Omega}}_{(i)}$, and 5 repetitions for RF using either random sampling or active learning on the dimension-reduced version of the pressure $P^*$. For each combination, we take the mean accuracy at each training set size with a 1$\sigma$ deviation band. 
\section{Results}

This section is divided into two parts -- the first addresses the development and selection of an adequate machine learning model for the EoS classification problem, as well as the performance of active vs. traditional learning implementations. Secondly, we discuss the deployment of the best-performing model, what is learned about the correspondence between EoS parameter space and stability classes, and implications for the modeling of heavy-ion collisions and experimental searches for the QCD critical point. 


\begin{figure}
\center
\includegraphics[width=0.8\linewidth]{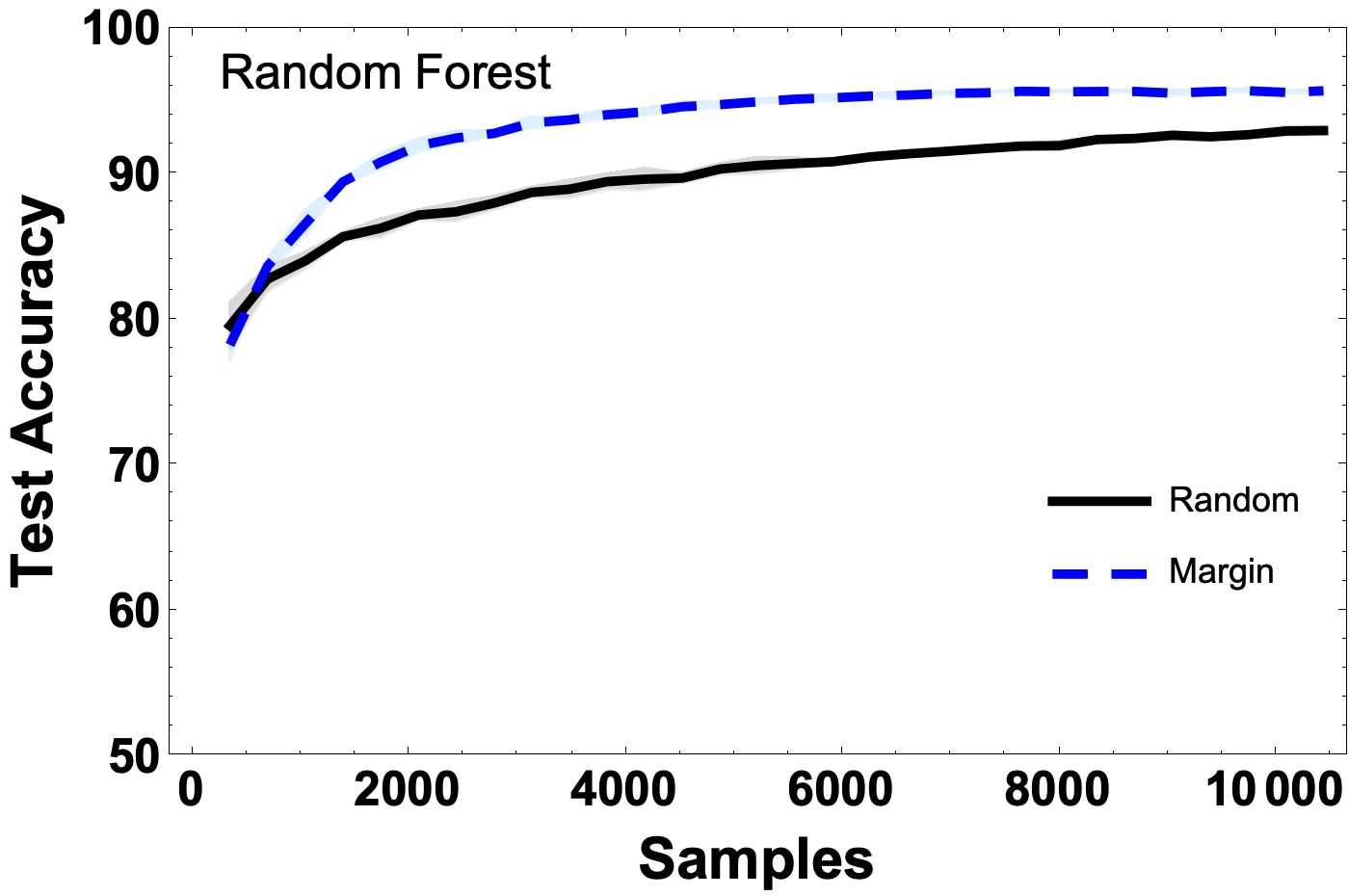}
\includegraphics[width=0.8\linewidth]{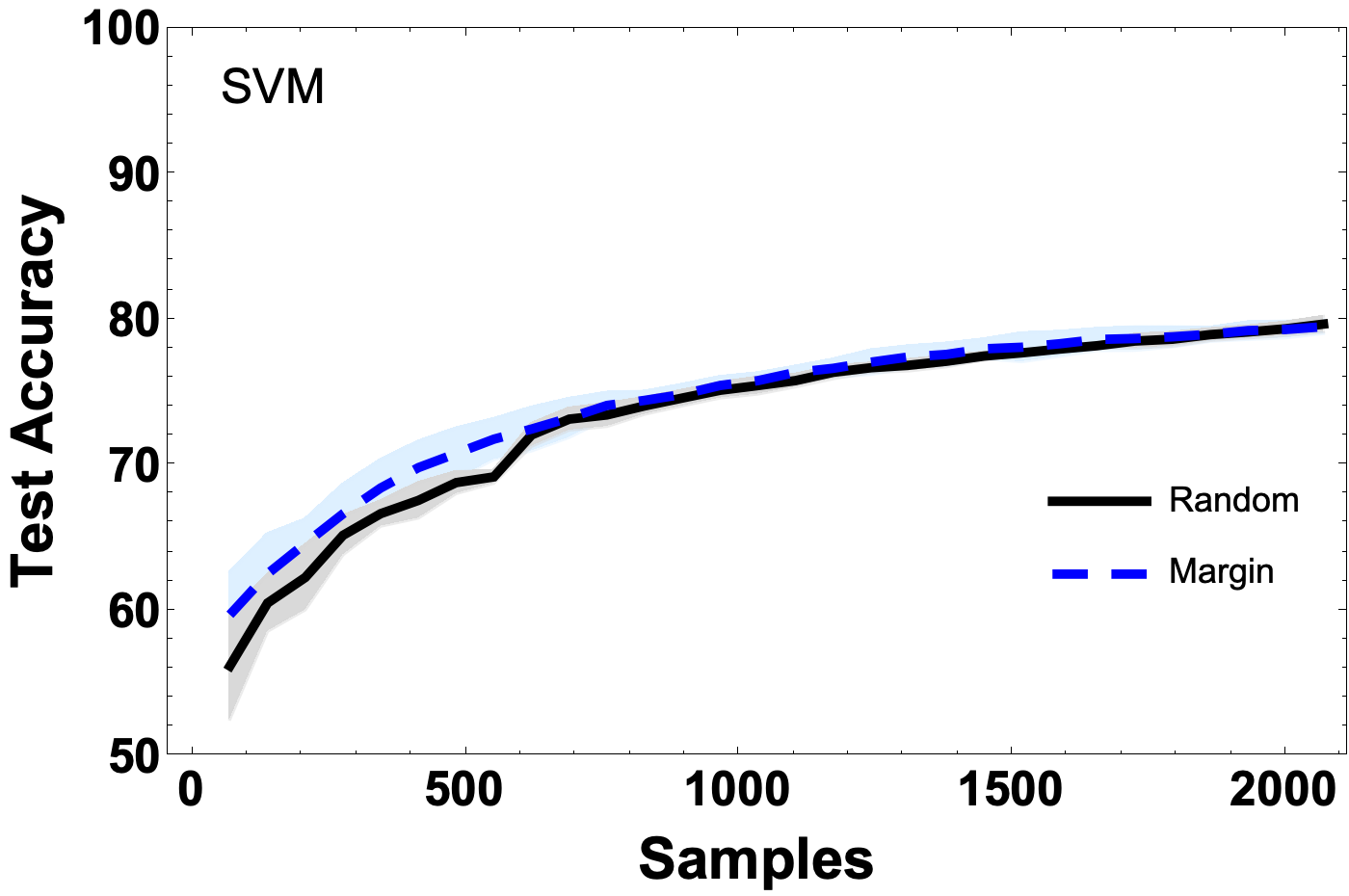} 
\includegraphics[width=0.8\linewidth]{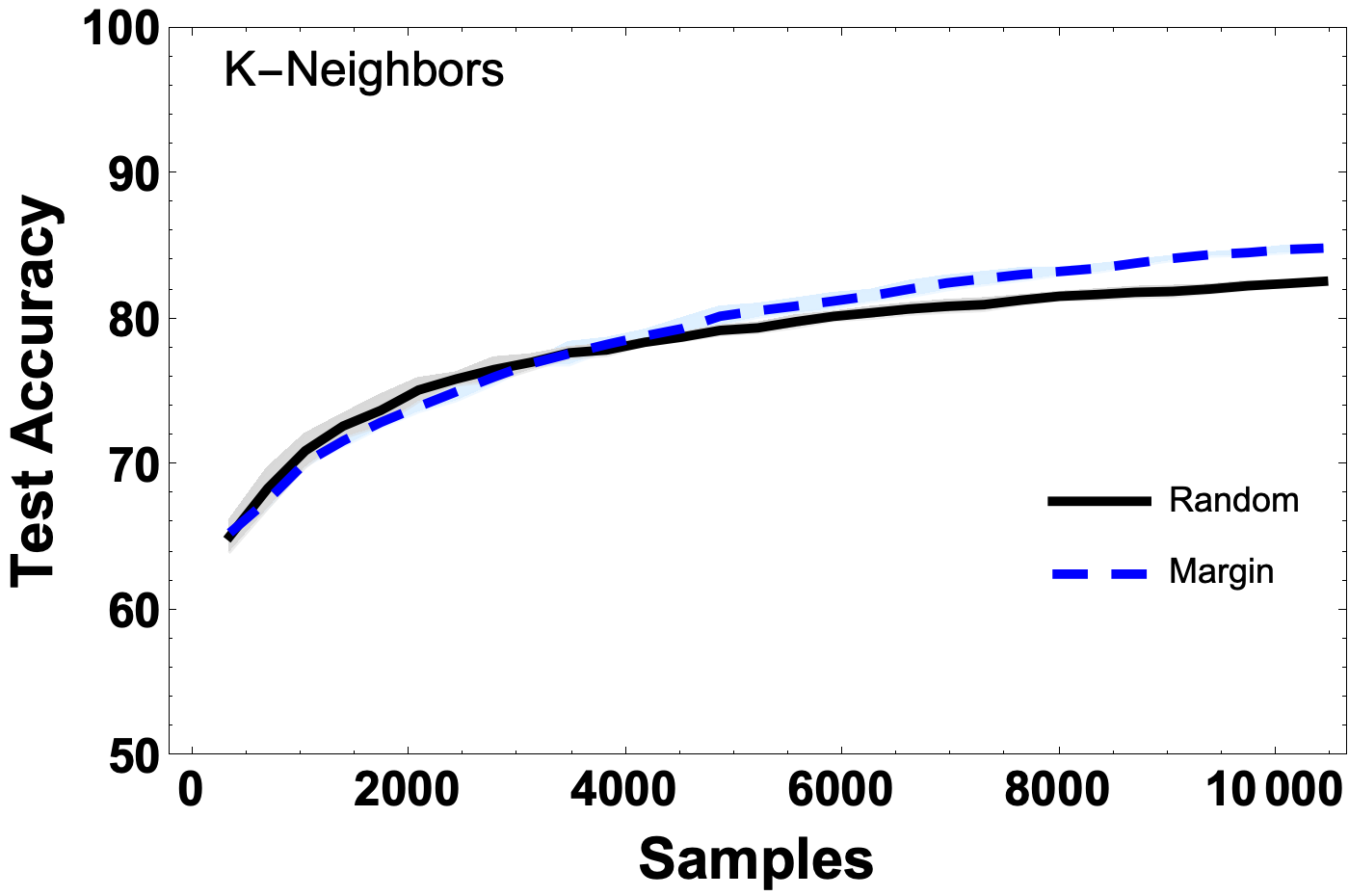}
\caption{Average performance on the test set as a function of training set size. The black lines correspond to the performance of models trained using random sampling and the dashed blue lines correspond to models trained using active learning. The bands show 1$\sigma$ deviations from the average.} 
\label{fig:avg_paramonly}
\end{figure}

\subsection{Model development}

The primary aspect of developing a machine learning model is to track how performance evolves during training. Fig.~\ref{fig:avg_paramonly} shows test accuracy as a function of training set size for each class of models trained on only the input $\vec{\Omega}_{(i)}$ data (recall that we distinguish this from training directly on the EoS itself that is signified by $P^*$). The solid and dashed lines represent the average behavior for a class of models across 5 runs with random sampling and active learning, respectively, and a corresponding 1$\sigma$ uncertainty band. Generally, a performance measure, which in this case is the recorded test set accuracy, is expected to improve on average as the number of training samples increases. We also re-emphasize that, during training, the model is completely blind to the test set and there is no overlap between test and training points. From Fig.~\ref{fig:avg_paramonly}, it is clear that active learning provides a significant advantage for RF and KNN models. In the SVM case, there is an initial advantage that vanishes when the training set reaches a size of $\sim$700 samples. 

We are interested in the combination of input data, sampling, and learning algorithm that performs best given the constraints set for label acquisition. From Fig.~\ref{fig:avg_paramonly}, we see that RF coupled to active learning clearly outperforms other models, with test set accuracy quickly converging around 96\% and exhibiting small variability. This agrees with our understanding of the benefit behind ensemble methods, where voting over multiple (nearly uncorrelated) models tends to reduce the bias and the variance component of the error significantly, see for example Refs.~\cite{Hastie09,Murphy2012,Bishop2006,Goodfellow2016,Mehta2019} for an in-depth discussion on the bias-variance trade-off in machine learning. 

Although a consistent final accuracy of 96\% is good enough for most applications of learning algorithms, we wanted to investigate if an even higher accuracy could be achieved using the dimension-reduced pressure $P^*$ as the training data. Because generating $P^*$ and training models using it as input is significantly more computationally expensive than the previous case, we limit this analysis to RF algorithms. This is expected considering the complexity cost behind SVM and KNN (for small K). As an example, SVMs are mathematically represented by a convex optimization problem. Ensemble methods like random forests, gradient boosting, bagging and others \cite{Hastie09,geron2022} normally use shallow decision trees as weak learners. Such decision trees are easy to train and require normally less CPU cycles than convex optimization problems like SVMs, which involve affine transformations with dense matrices. 

\begin{figure}
\centering
\begin{tabular}{c}
\includegraphics[width=0.8\linewidth]{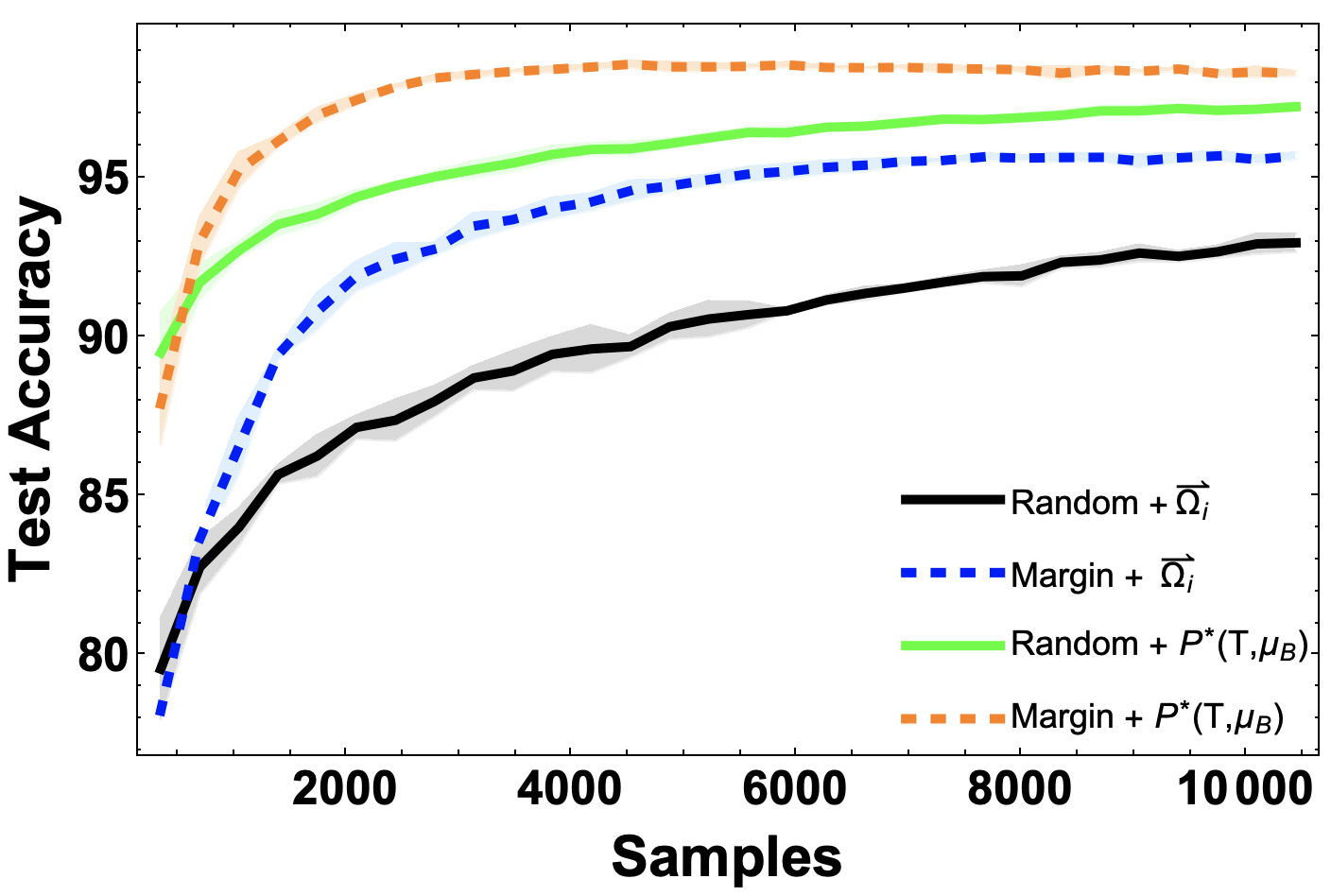} \\
\end{tabular}
\caption{Accuracy on the test set for RF models trained on input parameters coupled to random (solid black) and margin (dashed blue) sampling methods. This is compared to RF models trained on the transformed pressure coupled to random (solid green) and margin (dashed orange) sampling methods.} 
\label{fig:paramonly_vs_fullpress}
\end{figure}

We compare the performance of RF models using different training data in Fig.~\ref{fig:paramonly_vs_fullpress}, which shows the accuracy on the test set as a function of the training set size during training. The dashed orange and solid green lines correspond to margin and random sampling, respectively, using $P^*$ as input. The dashed blue and solid black lines correspond to margin and random sampling, respectively, using $\vec{\Omega}_{(i)}$ as input. The matching bands represent a 1$\sigma$ deviation from the mean performance over 5 runs. For both input classes, active learning outperforms random sampling. Surprisingly, when RF algorithms are trained using $P^*$, they outperform RF models trained on $\vec{\Omega}_{(i)}$ using active learning in all stages of training. The classifier performs even better when trained on $P^*$ with active learning, consistently achieving nearly perfect accuracy with under 3,000 samples in the training set. 

This increase in accuracy is likely due to the nature of the map between the different input spaces and the EoS classes. The map between $\vec{\Omega}_{(i)}$ and the stability classes is highly non-linear, whereas in $P^*$ space, the transition between classes is likely simpler to model in terms of input variables. Regardless, a strong classifier can be achieved with either set of input data. Using $P^*$ sacrifices the computational advantage over non-ML classification for near perfect classification accuracy, while models trained on $\vec{\Omega}_{(i)}$ peak at slightly lower accuracy, but with a significantly lower computational cost. We discuss execution-time benchmarking in detail in Sec.~\ref{time}. 

\begin{figure}
\centering
\begin{tabular}{c}
\includegraphics[width=0.8\linewidth]{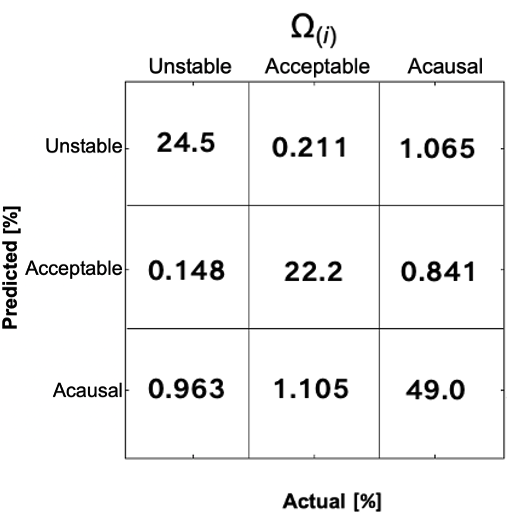} \\
\includegraphics[width=0.8\linewidth]{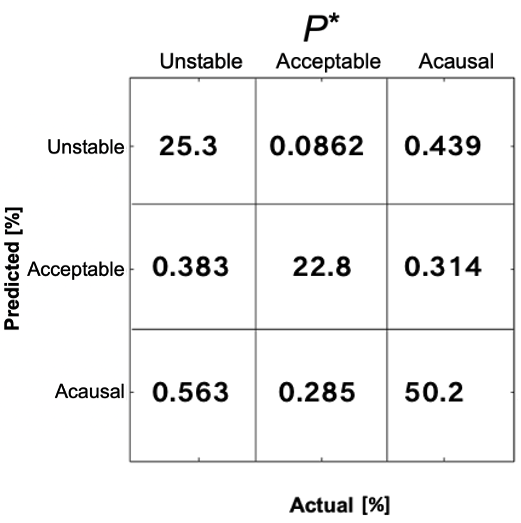}
\end{tabular}
\caption{Confusion matrix averaged over the final random forests models' performance on the test set, after training with $\Omega_{(i)}$ (top) or $P^*$ (bottom).} 
\label{fig:confusionmatrix}
\end{figure}

The summary of predictions on the test set is given by the confusion matrix \cite{sammut2010} of the model. In Fig.\ \ref{fig:confusionmatrix}, we show the confusion matrix for both classes of models with active learning, where the columns represent the true class (as calculated thermodynamically), and the rows indicate the class predicted by the model. We normalize the number of points in each entry by the total number of points in the test set, and show the corresponding percentage. The diagonal elements are the percentage of points that belonged to a certain class and were classified correctly by the model. Correspondingly, the off-diagonal elements quantify the percentage of points in the test set that were misclassified by the model. 

From Fig \ref{fig:confusionmatrix}, we see that acausal EoS are the most problematic class for models trained on $\vec{\Omega}_{(i)}$. Out of the average 4.33\% of points that were misclassified, an average total of 2.07\% were incorrectly predicted to be acausal, while another average 1.906\% of points that were acausal ended up misclassified as either acceptable or unstable. Combined, acceptable/unstable points being incorrectly classified as acausal and, on the other hand, acausal points being incorrectly classified as acceptable/unstable account for 92\% of misclassifications on average. If we break down the confusion between acausal and acceptable/unstable individually, we see that most incorrectly classified acausal points are, in reality, unstable, and vice-versa. 

In the context of the EoS parameter space, this means that there is a clear distinction between acceptable and acausal/unstable EoS, but the boundary between acausal and unstable can become fuzzy in certain regions. In this aspect, using $P^*$ as training input seems to help, but not significantly. As shown in Fig \ref{fig:confusionmatrix}, out of the average 2.07\% of points that were misclassified, confusion in acausal classifications accounts for 77.4\% of the mistakes.

In practice, the most important aspect of the confusion matrix analysis is to evaluate the prevalence of false positives/negatives. A false positive would be an EoS that is unstable/acausal, but gets incorrectly classified as acceptable. A false negative is an acceptable EoS that gets incorrectly classified as unstable/acausal. From Fig.\ \ref{fig:confusionmatrix}, we see that the false negative and positive rates for the models trained on $\vec{\Omega}_{(i)}$ are on average 1.316\% and 0.989\%, respectively. In the $P^*$ cases, the average rates are 0.371\% for false negatives and 0.697\% for false positives. 

The incidence of false positives/negatives for the class of models trained on $P^*$ is about half of those trained on the input vectors, but in either case these rates are low enough for most applications. In general, models taking in $P^*$ are more suitable for analyses that require knowledge of a specific point, since the overall accuracy is higher. However, models trained on $\vec{\Omega}_{(i)}$ still provide an accurate description of the EoS parameter space and class structure.



\subsection{Model Deployment}\label{deployment}

In addition to training and comparing the performance of different classes of models, we illustrate a deployment framework by analyzing the features of the EoS parameter space relevant to experimental searches for the critical point. The analysis was done using the top performing model in terms of classification accuracy and execution time, namely random forests trained with active sampling and input vectors $\vec{\Omega}_{(i)}$. From here on, we will refer to this model as RF$^A_\Omega$, where RF stands for random forests, A for active learning, and $\Omega$ refers to the type of training data. 

\subsubsection{Execution time benchmarking}\label{time}

\begin{figure}
\centering
\includegraphics[width=0.8\linewidth]{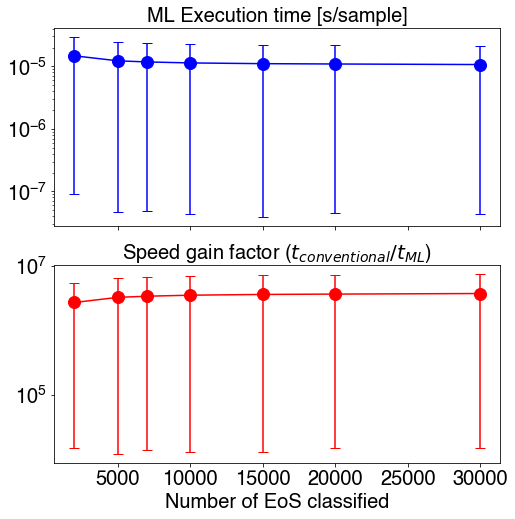} \\
\caption{Top panel: execution time per sample on a log scale as a function of the total number of samples with ML classification. Bottom panel: speed factor gained using ML versus conventional classification. In both cases the error bars represent a 68\% confidence interval based on jackknife resampling of 50 samples.} 
\label{fig:timing}
\end{figure}

One of the important metrics when choosing a model for deployment is accuracy. The RF$^\textrm{A}_\Omega$ model yields a final test set accuracy of 96.772 \%. This is not as high as the models that were trained on $P^*$. However, it is important to also keep track of the execution time. Machine learning assisted classification, if implemented appropriately, should yield a significant computational advantage. Without ML assistance, the classification of EoS stability is $\mathcal{O}(1-2)$ in seconds. The execution time of ML-assisted classification can be calculated as a per sample rate, classifying a certain number of samples in bulk, then dividing the total execution time by the total number of samples. This is shown in the top panel of Fig. \ref{fig:timing}, which displays the execution time in seconds per sample for RF$^\textrm{A}_\Omega$ classification as a function of the number of EoS classified. In order to test the robustness of the model, we repeat this rate calculation 50 times and show the 68\% confidence interval based on jackknife resampling. The model consistently performs at the microsecond scale. 

In the bottom panel of Fig. \ref{fig:timing}, we show the speed gain factor as a function of the number of EoS classified. This was calculated by dividing the ML execution time per sample by the non-ML classification time per sample. The same statistical methods were used for constructing the confidence intervals. RF$^\textrm{A}_\Omega$ assisted classification consistently provides a computational advantage of 5-6 orders of magnitude. 

\subsubsection{EoS Stability Analysis}\label{model_results}
The speed and high accuracy of RF$^\textrm{A}_\Omega$ allow us to map the stability of the EoS as a function of input parameters in fine detail. These parameters relate to key physical properties of the QCD critical point. As discussed Sec.~\ref{EoS}, $\alpha_\textrm{diff}$ represents the angular separation between Ising axes $(r,h)$ in the mapping to QCD variables, $w$ globally scales the Ising axes  -- i.e. the critical region --  and $\rho$ stretches the critical region along the transition line ($\mu_B$ direction). 

\begin{figure}
\centering
\includegraphics[width=\linewidth]{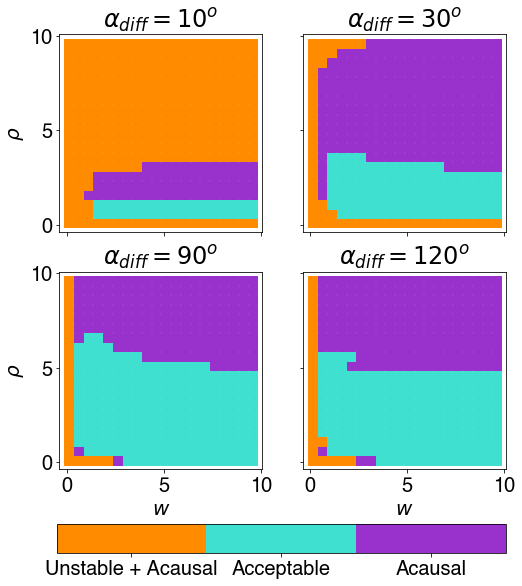}
\caption{Stability and causality regions on the $w-\rho$ plane for different $\alpha_\textrm{diff}$ values at fixed $\mu_{BC} = 400$ MeV. As the angular parameter moves away from orthogonality, the stability region shrinks. An upper limit appears for $\rho$, which drives EoS to acausal regions when too large.} 
\label{fig:param_space}
\end{figure}

 In Fig. \ref{fig:param_space}, we fix $\mu_B = 400$ MeV and determine the stability/causality of the EoS for different values of the angular parameter $\alpha_\textrm{diff}$ as a function of the scaling parameters $w$ and $\rho$ on a grid much finer than previous calculations \cite{Parotto:2018pwx}. From Fig. \ref{fig:param_space} it can be inferred that the stability region in the $w-\rho$ plane shrinks as the angular parameter moves away from orthogonality, and that there is a hard limit on the value of $\rho$, which drives the EoS to acausal regimes when too large. Hence, under the current mapping, the critical region cannot be too large in the $\mu_B$ direction.
 
  The exact value of the $\rho$ stability cut-off depends strongly on the value of $\alpha_\textrm{diff}$, but not as much on the global scaling of the critical region determined by $w$. These results reflect the underlying physics of the model \cite{Mroczek:2020rpm}. The EoS with critical contributions is matched to lattice QCD data by construction, and these parameters can stretch the critical region along the transition line to the point where the EoS cannot be simultaneously acceptable and consistent with lattice QCD. If either $\alpha_\textrm{diff}$ or $\rho$ spreads the critical region too broadly across $\mu_B$, the EoS will become acausal. Stability and causality also depend on $\mu_{BC}$, which is discussed below and in Section~\ref{sec:correlations}.

\begin{figure}
\centering
\includegraphics[width=\linewidth]{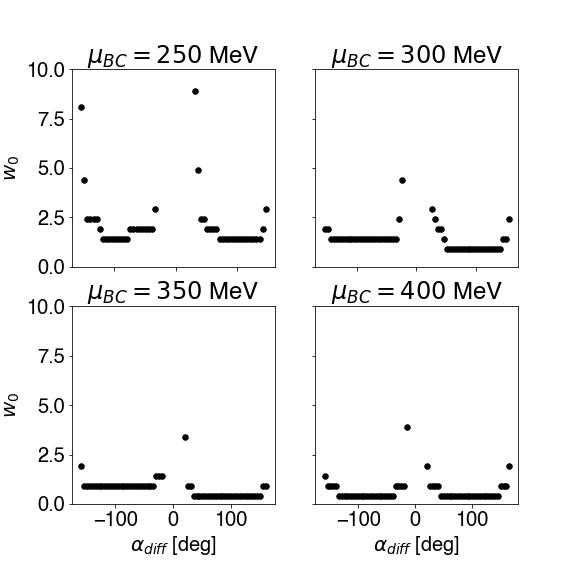}
\caption{Smallest possible value $w_0$ (largest global scaling of the critical region thermodynamically allowed) for different values of $\mu_{BC}$ and $\rho = 2$ as a function of $\alpha_\textrm{diff}$.} 
\label{fig:biggestCP}
\end{figure}

Another point of interest is to maximize the overall size of the critical region.
This is reflected by the minimum value of $w$, to which to the overall size of the critical region is (mildly) inversely proportional. In general, one has $\Delta \mu_B \Delta T \sim w^{-2/7}$ \cite{Mroczek:2020rpm}, where $\Delta \mu_B, T$  are the corresponding sizes of the critical region in the $T$ and $\mu_{BC}$ directions. Hence, despite the overall size of the critical region only being midly affected by $w$, it determines the largest possible critical region for a particular choice of $\mu_{BC}$, $\rho$, and $\alpha_\textrm{diff}$. This analysis is shown in Fig. \ref{fig:biggestCP}, which displays the smallest acceptable value $w=w_0$ for different values of $\mu_{BC}$ and fixed $\rho = 2$ as a function of $\alpha_\textrm{diff}$. We see that, as $\alpha_\textrm{diff}$ moves away from $\pm$ 90$^\textrm{o}
$ (orthogonal Ising axes), it drives the value of $w_0$ up. In addition, as $\alpha_\textrm{diff} \rightarrow 0$, stability disappears entirely. The band where no EoS are possible shrinks as we move $\mu_{BC}$ to larger values, as would be expected since large $\mu_{BC}$ interferes the least with lattice results at $\mu_B=0$. Most importantly, we see that regions of low $w_0$ appear at values of $\alpha_\textrm{diff}$ closer to $90^\textrm{o}$, meaning that these regimes are compatible with a larger critical region. When $\mu_{BC}$ is larger, the low $w_0$ regions extend for longer in $\alpha_\textrm{diff}$, because moving the CP away from $\mu_B =0$ allows for a larger CP to still be consistent with the matching to lattice QCD.  To summarize, we find that placing the critical point at larger $\mu_{BC}$ guarantees the most flexibility in the possible size and shape of the critical region. However, even when $\mu_{BC}$ is large, the Ising axes cannot come too close together without causing pathological behavior.

\subsection{Correlations between input parameters in acceptable EoS}\label{sec:correlations}

In the process of developing and training the ML models presented in this work, nearly 40k realizations of the BEST EoS were labeled. The pool of labeled EoS includes samples taken randomly and with active learning. Since active queries oversample along the boundary, this collection of EoS should strongly reflect the true stable and causal regions. By selecting only the acceptable EoS from this pool, we can gain insight on the distribution and correlations between input values in the stability/causality windows. 
The analysis of the acceptable training samples is presented in this work as a tool complementary to the ML models because the training data reflects the true distributions and correlations between parameters for acceptable EoS. These samples should be studied because they display the trends ML classification should follow, aside from providing a general intuition for the regime of thermodynamic validity of the EoS model. 

\begin{figure*}[htb]
\includegraphics[width=0.8\textwidth]{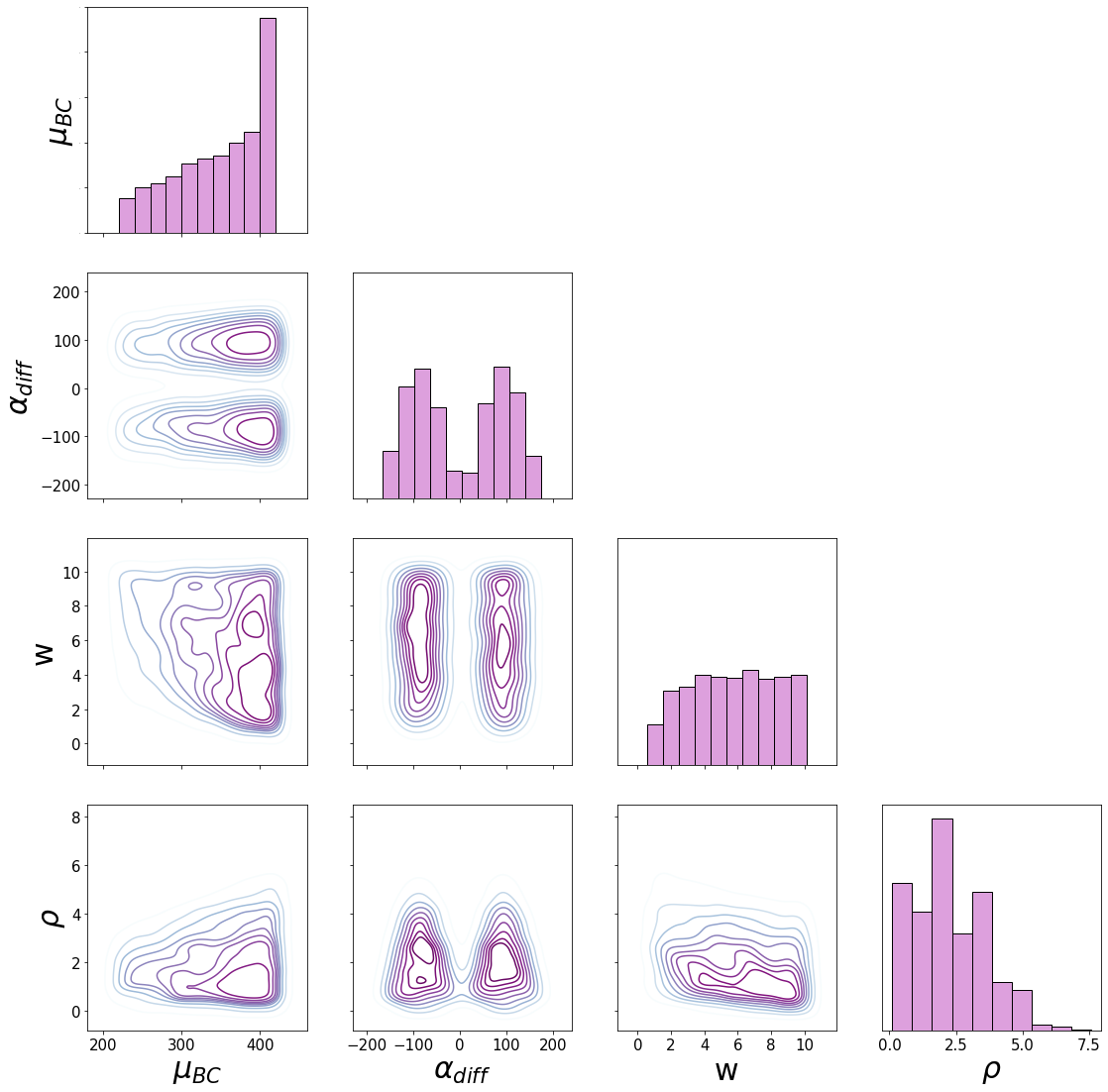}
\caption{Off-diagonal: pair-wise correlations between input parameters for acceptable EoS in the combined training set. Each plot shows the density distribution for pairs of input parameters. Diagonal: Histograms for each parameter and its distribution in the class of acceptable EoS.}
\label{fig:pair_wise}
\end{figure*}

The histogram for each input parameter is shown along the diagonal in the corner plot in Fig.~\ref{fig:pair_wise}. We see acceptable EoS are more likely to stem from larger values of $\mu_{BC}$, values of $\alpha_{\textrm{diff}}$ close to $90^\textrm{o}$, and lower $\rho$. There is not a strong dependence on $w$, but the number of acceptable EoS decreases for $w \lesssim 1$. The observations are in line with general arguments on the size and shape of the critical region \cite{Mroczek:2020rpm}. The peak at $\rho\approx 2.0$ is likely due to active learning, since this seems to be the point where EoS in the intermediate angle regime (10$^\textrm{o}$ $\leq \alpha_{\textrm{diff}} \leq$ 60$^\textrm{o}$) become acausal. These findings are consistent with what was found with ML-assisted classification using RF$^A_\Omega$. 

The off-diagonal elements of Fig.~\ref{fig:pair_wise} contain the pair-wise density correlations between input parameters. As expected, $\alpha_{\textrm{diff}}$ correlates strongly with other input parameters -- a CP further away from $\mu_B = 0$ and smaller $\rho$ allow for smaller angles. However, $\alpha_{\textrm{diff}}$ is always above $5^\textrm{o}$, even when $\rho$ is small, and $\rho$ is always below 5.0-6.0, even when  $\alpha_{\textrm{diff}}$ is not small. Thus, there is a limit for $\alpha_{\textrm{diff}}$ and $\rho$ in the current implementation of this model, since we cannot place the critical point at a larger value than $\mu_{BC}=420$ MeV due to limitations from lattice QCD. There is not a strong correlation between stability and $w$, because it is always possible to make the critical region small enough to suppress unstable behavior. Furthermore, we observe that larger critical regions ($w \lessapprox 1$) only appear for $\mu_{BC} > 300 $ MeV. This confirms the trends found by RF$^A_\Omega$ for the subsets of the parameter space discussed in Section~\ref{model_results}. 

In summary, $\mu_{BC} \gtrapprox 300 $ MeV provides the most freedom, but it always holds that $\rho \lessapprox 5.0$ and $\alpha_{\textrm{diff}} \gtrapprox 5.0^\textrm{o}$ under the current mapping. Recall that any pathological behavior of the EoS is due to tension with lattice calculations. This is model-dependent because of the choice of mapping between Ising and QCD variables and the truncation of the lattice Taylor expansion. Changing either would affect the quantitative results discussed in this work, which are specific to the EoS presented in Ref.~\cite{Parotto:2018pwx}.

\section{Conclusions}
The BEST collaboration EoS relies on a non-universal linear map of 3D Ising model variables onto the QCD phase diagram, which contains four free parameters. A subset of the resulting four-dimensional parameter space leads to unstable and/or acausal realizations of the EoS. In this work, we built a machine learning framework that incorporates active learning to guide the model towards the most important regions in the input parameter space; this helps to efficiently rule out unphysical EoS with high accuracy. In addition to mapping the stability and causality of the EoS as a function of the input parameters across the entire available parameter space, we find that certain mapping parameters are constrained to $\alpha_{\textrm{diff}}\gtrapprox 5^\textrm{o}$ and $\rho \gtrapprox 5$. Additionally, a strong preference for a critical point at large baryon chemical potentials $\mu_{BC}$ is shown, especially when the critical region is large. Low $\mu_{BC}$ can only coexist with significantly smaller critical regions. Although these findings are quantitatively strongly dependent on the actual implementation of the BEST EoS, their qualitative nature is likely to be quite general, as it essentially stems from the (in)compatibility of the universal critical behavior and first-principle lattice QCD results. 

The insights presented in this work can be used in future hydrodynamic studies of the evolution of matter created in ultra-relativistic heavy-ion collision experiments at low beam energies. Currently for heavy-ion collisions, it is not possible to directly compare the EoS to experimental data.  Instead, one must run relativistic viscous hydrodynamic simulations with a large number of free parameters that are then directly compared to experimental data. The free parameters are constrained using a combination of emulators and Bayesian analysis \cite{Pratt:2015zsa,Bernhard:2016tnd,Bernhard:2019bmu,JETSCAPE:2020mzn,Nijs:2020roc}, which are limited by the enormous amount of computational time required to run a single parameter set. The results presented here significantly cut down the input parameter space, allowing for tighter priors in a potential Bayesian analysis comparing heavy-ion hydrodynamics simulations to experimental data.


This is the first time that active learning has been employed in the context of heavy-ion collisions. We demonstrated that active learning can significantly reduce sampling requirements for training classifiers to search for acceptable EoS. Because of the speed and accuracy we reached in our framework using active learning, our methodology promises to be useful for a number of problems in the field of heavy-ion collisions. Additionally, the machine learning pipeline developed in this work is generic enough that it can be applied to any EoS with a parameter-space-to-class correspondence.

\begin{acknowledgments}

D.M is supported by the National Science Foundation Graduate Research Fellowship Program under Grant No. DGE – 1746047, the Illinois Center for Advanced Studies of the Universe Graduate Fellowship, and the University of Illinois Graduate College Distinguished Fellowship. The work of MHJ is supported by the U.S. Department of Energy, Office of Science, office of Nuclear Physics under grant No. DE-SC0021152 and U.S. National Science Foundation Grants No. PHY-1404159 and PHY-2013047. J.N.H. acknowledges the
support from the US-DOE Nuclear Science Grant No. DE-SC0020633. This material is based upon work supported by the National Science Foundation under grants no. PHY-1654219 and PHY-2116686. 
This work was supported in part by the National Science Foundation within the framework of the MUSES collaboration, under grant number OAC-2103680.
The authors also acknowledge support
from the Illinois Campus Cluster, a computing resource that
is operated by the Illinois Campus Cluster Program (ICCP)
in conjunction with the National Center for Supercomputing
Applications (NCSA), and which is supported by funds from
the University of Illinois at Urbana-Champaign. 
This work was completed in part with resources provided by the Research Computing Data Core at the University of Houston.
\end{acknowledgments}

\bibliographystyle{unsrt}
\bibliography{apssamp,inspire}

\end{document}